\documentclass{PoS}

\usepackage{epsfig}
\newcommand{\bra}[1]{\left< #1 \right|}
\newcommand{\ket}[1]{\left| #1 \right>}

\title{The Boer-Mulders Function: Models and Universality}
\ShortTitle{The Boer-Mulders Function: Models and Universality}
\author{\speaker{Brian Hannafious}\\
       New Mexico State University, USA\\
       E-mail: \email{brianh@nmsu.edu}}
\author{Matthias Burkardt\\
        New Mexico State University, USA\\
        E-mail: \email{burkardt@nmsu.edu}}
\abstract{
The sign of the Boer-Mulders function $h_1^\perp$ is related to the sign of the GPD $\bar E_T$ through the mechanism
of chromodynamic lensing. Model calculations of the sign of $\bar E_T$
indicate that the sign of $h_1^\perp$ may be the same in all ground state hadrons.}
\FullConference{LIGHT CONE 2008 Relativistic Nuclear and Particle Physics\\
                July 7-11, 2008\\
                Mulhouse, France}
\begin{document}

\section{The Boer-Mulders Function and $\bar E_T$}
In momentum space, the distribution of polarized quarks in an unpolarized target is given by the expression
\begin{equation}
	f_{q^\uparrow/p} = \frac{1}{2}
	\left[
		f_1^q(x,k^2_T) - h_1^{\perp q}(x,k^2_T)
		\frac{\left(\hat{\mathbf P} \times \mathbf k_T\right) \cdot \mathbf S_q}{M}
	\right],
\end{equation}
where $S_q$ is the quark spin, and $h_1^\perp(x,k^2_T),$ the Boer-Mulders function, describes a momentum space
asymmetry. In the Trento conventions \cite{Bacchetta:2004jz}, for a target approaching the observer and a
positive $h_1^\perp(x,k^2_T),$ spin up quarks preferentially move towards the left.

A similar expression can be written down in position space by making use of impact parameter dependent 
parton distributions,
\begin{equation}
	F^q_{\mathbf s}(x,\mathbf b) = \frac{1}{2}
	\left[
		H(x,\mathbf b^2) - S_q^ib^j\epsilon^{ij}\frac{1}{m}
		\left(E_T^\prime(x,\mathbf b^2) + 2\widetilde H_T^\prime(x,\mathbf b^2)\right)
	\right].
\end{equation}
It is convenient to define the quantity
$\bar E_T(x,\mathbf b^2) = E_T^\prime(x,\mathbf b^2) + 2\widetilde H_T^\prime(x,\mathbf b^2),$
which describes a sideways shift in the position of polarized quarks in an unpolarized hadron \cite{Diehl:2005jf}.
While the Boer-Mulders function requires a final state interaction to exist, $\bar E_T(x,\mathbf b^2)$ is an
intrinsic property of hadrons. However, $\bar E_T(x,\mathbf b^2)$ is the position space analogue of
$h_1^\perp(x,k^2_T)$ in the sense that the signs of the functions are negatively correlated through the
mechanism of chromodynamic lensing \cite{Burkardt:2005hp}, which transforms position space asymmetries into
momentum space asymmetries through attractive final state interactions.

Model calculations indicate that the sign of the Boer-Mulders function is likely the same in all ground state hadrons
\cite{Burkardt:2007xm}. In order to explore this, one would like to perform model calculations of the sign of
$h_1^\perp(x,k^2_T).$ However, it is often more straightforward to calculate the sign of $\bar E_T(x,\mathbf b^2)$
in position space, and then employ chromodynamic lensing to infer the sign of $h_1^\perp(x,k^2_T).$

\section{$\bar E_T(x,\mathbf b^2)$ in the Bag model}
As a general Bag model wave function, take the Dirac spinor
\begin{equation}
\label{spinor}
\Psi_m = 
\left(
 \begin{matrix}
  if \chi_m \\
  -g (\vec{\sigma} \cdot \hat{x}) \chi_m
 \end{matrix}
\right),
\end{equation}
where $f$ is a monotonically decreasing radial function, $g$ is the derivative of $f,$ as required by the free
Dirac equation, and $\chi_m$ is a Pauli spinor.

The impact parameter dependent parton distributions that we would like to evaluate are of the form
\begin{equation}
F_\Gamma (x,{\bf b}_\perp ) =
{\cal N}^{-1} \int \frac{dz^-}{4\pi} e^{ixp^+z^-} 
\left\langle p^+ , {\bf 0}_\perp \right| \bar{q}(0,{\bf b}_\perp)
\Gamma {q}(z^-,{\bf b}_\perp)\left| p^+, {\bf 0}_\perp
\right\rangle.
\end{equation}
Complications arising from computing light-like correlation functions in the Bag model can be avoided by studying
the lowest moment of the GPDs, 
\begin{equation}
\int dx F_\Gamma (x,{\bf b}_\perp) = const. \int dx^3
 \left\langle  {\bf {\vec 0}}\right| \bar{q}(x^3,{\bf b}_\perp)
\Gamma {q}(x^3,{\bf b}_\perp)\left| {\bf {\vec 0}}
\right\rangle,
\label{foo0}
\end{equation}
Translational invariance has been used to localize the states in Eq. (\ref{foo0}) to the origin.

Quarks with transverse polarization ${\bf s}$ are projected out by the operator
$\frac{1}{2}\bar{q}\left[\gamma^+ -s^ji\sigma^{+j}\gamma_5\right]q$ \cite{Diehl:2005jf} and therefore the vector field
representing the transverse quark polarization density is given by $-i\bar{q}\sigma^{+j}\gamma_5 q$. We thus consider impact
parameter dependent PDFs with $\Gamma = -i \sigma^{+j}\gamma_5,$ which are related to the Fourier transforms of the chirally odd
GPDs $\bar{E}_T$, $H_T$ and $\tilde{H}_T$ \cite{Diehl:2005jf} 
\begin{equation}
F_T^i= -\varepsilon^{ij}b^j\frac{1}{M}{\bar{\cal E}}_T' 
+ S^i\left({\cal H}_T - \frac{1}{4M^2}\Delta_b \tilde{\cal H}_T
\right) + \left(2b^ib^j-b^2\delta^{ij}\right) S^j
\frac{1}{M^2}\tilde{\cal H}_T'',
\end{equation}
where script letters denote the Fourier transforms of the GPDs, and $S^j$ is the spin of the target. Only the term involving
$\bar{\cal E}_T$ contributes for an unpolarized target, which is why it is only $\bar{E}_T$ that is expected to be related
to the Boer-Mulders function. The term can be extracted by considering the density corresponding to $\Gamma=-i\sigma^{+j}\gamma_5$
and summing over the target spin. For a single quark state this procedure yields
\begin{eqnarray}
 \sum_m \bra{PS_{m}}\bar{\Psi}(x^3,\mathbf{b}_\perp)
	i\sigma^{+i}\gamma_5\Psi(x^3,\mathbf{b}_\perp) \ket{PS_{m}} \nonumber\\
 = \label{bm_exp} -\frac{1}{\sqrt{2}} \sum_m
	(f^2 + g^2) s^i_m + 2fg\epsilon^{ij}\hat{b}_\perp^j - 2g^2\hat{b}_\perp^i
		(\hat{b}_\perp\cdot \vec{s}_m) 
\label{eq:fg}
\end{eqnarray}
where $\vec{s}_m$ is the spin vector corresponding to the Pauli spinor $\chi_m$.  The first and last terms of (\ref{bm_exp})
do not survive the sum over `target' polarizations. The asymmetry is given entirely by the middle term, which is an interference
between the upper and lower components of Eq. (\ref{spinor}).  For the lowest moment of $\bar{\cal E}_T,$ we find
\begin{equation}
\kappa_T=
\int\! dx \bar{E}_T(x,0,0) =\int\! dx d^2{\bf b}_\perp 
\bar{\cal E}_T
	=
	\frac{2MG}{3\sqrt{2}\pi} \int_0^{R_0} dr \  r^3 fg.
	\label{res}
\end{equation}
The right hand side of (\ref{res}) is always positive because $f$ and $g$ are non-negative functions for $r$ less than the bag
radius, implying that $\bar{\cal E}_T\geq 0$.

In the bag model, the correlation between quark spin and quark orbital motion is the same, regardless of the orientation of $j_z$.
All quark spin orientations thus contribute coherently to $\bar{\cal E}_T^q$ and in the case of $d$ quarks, $\bar{\cal E}_T^d$ is 
equal to $\bar{\cal E}_T^d$ for a single quark, while for $u$ quarks it is twice as large. In fact, for any model where the quarks
are confined by some mean field potential one finds that all quark orbits give the same contribution to $\bar{\cal E}_T^q$ and
thus $\bar{\cal E}_T^q$ is equal to $\bar{\cal E}_T^q$ for a single quark orbit, multiplied by the number of quarks of flavor $q$.
In particular, in the large $N_C$ limit, where $N_u=N_d+1\rightarrow \infty$, the lowest $x$ moment of $\bar{\cal E}_T^q$ is the
same for $u$ and $d$ quark and both are of order ${\cal O}(N_C)$. Since the support of GPDs shrinks to $x = {\cal O}(1/N_C)$,
this implies that $\bar{E}_T^u(x,\xi,t)=\bar{E}_T^d(x,\xi,t)= {\cal O}(N_C^2).$

\begin{figure}
\epsfig{file=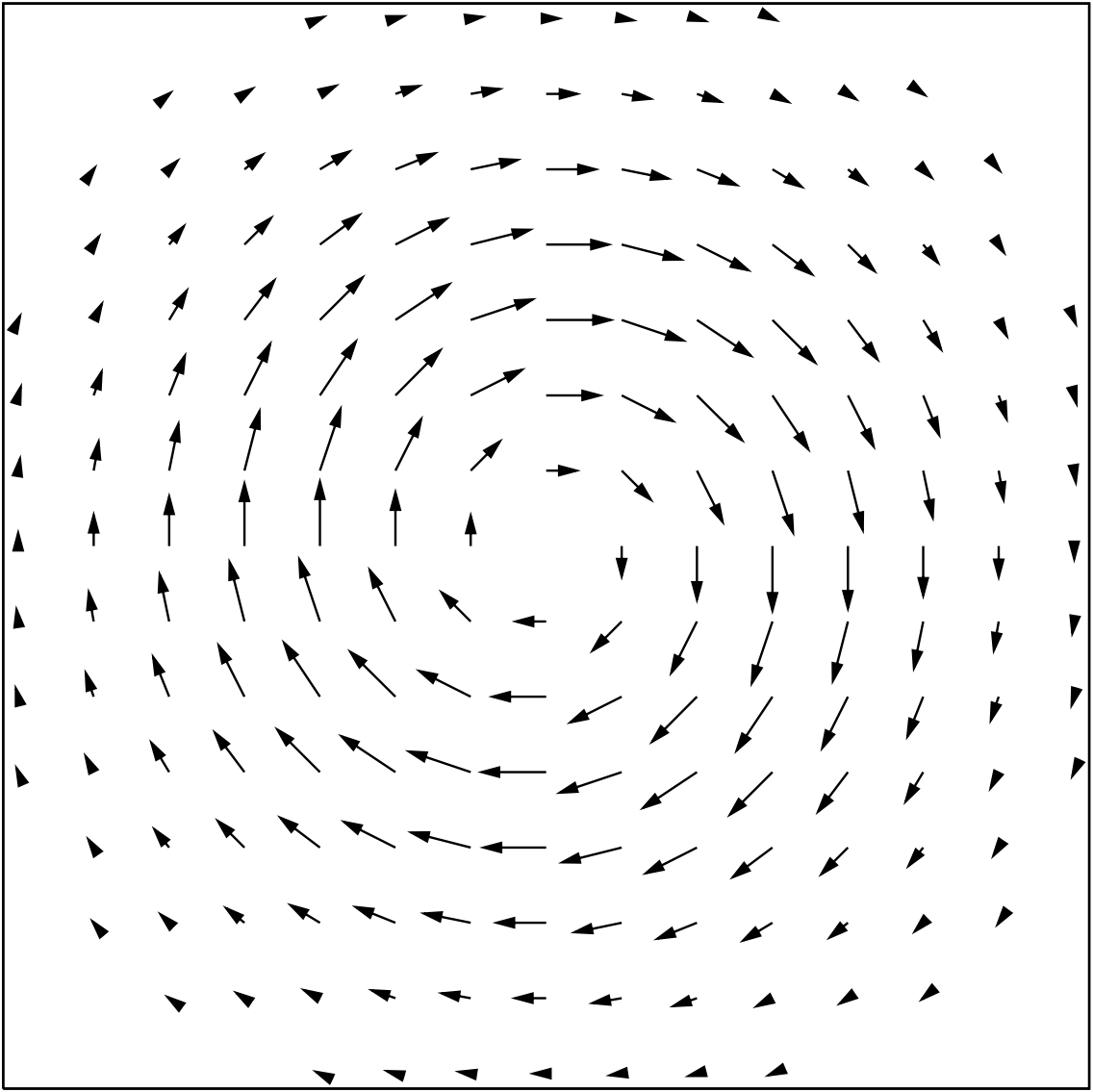, angle=0, width=\linewidth}
\caption{Lowest moment of the impact parameter dependent transversity distribution for an unpolarized
target in the MIT bag model. The `outside' of the spherical bag corresponds to the regions without arrows.}
\label{bag_vec_field}
\end{figure}

In order to visualise the transverse spin - impact parameter correlation in the bag model, the vector field 
\begin{equation}
-\int dx^3 fg\epsilon^{ij}\hat{b}^j
\label{foo1}
\end{equation}
representing the lowest moment of the transversity density in an unpolarized target has been plotted in Fig. \ref{bag_vec_field}
for bag model wave functions $f=j_0(r)$, and $g=j_1(r)$.

In the bag model, we thus obtain a counter-clockwise polarization for impact parameter dependent quark distributions, which
implies a negative Boer-Mulders function.

This result holds in potential models more general than the bag model, which has a scalar potential with the shape of an
infinite square well, and a vanishing vector potential. In the bag model, the upper and lower components of the Dirac equation,
$\phi_u$ and $\phi_l,$ satisfy
\begin{equation}
\phi_l = \frac{1}{E+m} {\vec \sigma}\cdot {\vec p} \phi_u.
\end{equation}
In the case of a general scalar potential, where the mass term $m(r)$ depends on the radius, and a general vector potential $V(r),$
this relationship becomes
\begin{equation}
\phi_l = \frac{1}{E+m(r)-V(r)} {\vec \sigma}\cdot {\vec p} \phi_u.
\label{pmdr}
\end{equation}
In order to avoid the Klein paradox, $V(r)$ cannot exceed $m(r),$ and so the denominator of Eq. (\ref{pmdr}) is positive.
Therefore, the results for the sign of the Boer-Mulders function are the same as in the bag model. In fact, the sign of the
spin-orbit correlation described by Eq. (\ref{eq:fg}) should be the same for the ground state of all
confining potential models.

The Boer-Mulders function has been calculated directly in the Diquark model in
\cite{Brodsky:2002rv,Burkardt:2003je,Gamberg:2003ey,Boer:2002ju,Bacchetta:2003rz,Radici:2007vc}, and $\bar E_T$ has been
directly calculated in the constituent quark model in \cite{Pasquini:2007xz}. Both calculations produce the same sign
for the Boer-Mulders function as the Bag model.
While these models involve interactions, they are contact interactions and the quarks mostly obey the free Dirac equation
that is responsible for the results from the Bag model.

Finally, the Bag model results also agree with the sign found on the lattice \cite{Gockeler:2006zu}.

\section{$\bar{E}_T$ in the Pion}
For the pion, the distribution of quarks with spin $s^i$ in impact parameter space reads
\begin{equation}
\frac{1}{2}\left[ F + s^i F_T^i \right] = 
 H(x,\mathbf{b}^2) + s^i \epsilon^{ij} b^j \frac{2}{m} \frac{\partial}{\partial \mathbf{b}^2} \bar{E}_T(x,\mathbf{b}^2),
 \label{E_pion}
\end{equation}
where ${\cal H}(x,\mathbf{b}^2)$ and $\bar{\cal E}_T(x,\mathbf{b}^2)$ are again the Fourier transforms of the GPDs $H(x,0,t)$ and
$\bar{E}_T(x,0,t)$ respectively. The definitions of $H(x,0,t)$ and $\bar{E}_T(x,0,t)$ whose definition are particularly simple,
\begin{eqnarray}
& & \int \frac{dz^-}{4\pi} e^{ixP^+z^-} \bra{\pi^\prime} \bar{q}(-\frac{1}{2}z) \gamma^+
   q(\frac{1}{2}z) q\ket{\pi} \vert_{z^+=0,z=0} 
= H(x, \xi, t) \nonumber \\
& & \int \frac{dz^-}{4\pi} e^{ixP^+z^-} \bra{\pi^\prime} \bar{q}(-\frac{1}{2}z)
   \sigma^{+j} \gamma_5 q(\frac{1}{2}z)\ket{\pi} \vert_{z^+=0,z=0} 
= \frac{1}{\Lambda} \bar{E}_T(x,\xi, t) \frac{\epsilon^{+j\alpha\beta}\Delta_\alpha P_\beta}{P^+}.
\nonumber\\
\end{eqnarray}
Here $\Lambda$ is some hadronic mass scale, which needs to be included in the definition if $\bar{E}_T(x,\xi, t)$ is
to be dimensionless.

Except for a slight change in the bag radius, the quark wave functions in the bag model are the same for pions and nucleons. 
Therefore, apart from a slight rescaling due to the different bag radii, $\bar{E}^u_T$ in a $\pi^+$ is the same as
$\frac{1}{2}\bar{E}^u_T$ or $\bar{E}^d_T$ in the proton. The factor $\frac{1}{2}$ accounts for the fact that there
are twice as many $u$ quarks in a proton as in a $\pi^+$. Most importantly, we find again the same sign for $\bar{E}_T$ as in 
the nucleon.

The Nambu-Jona-Lasino (NJL) model of the pion produces the same sign for the Boer-Mulders function as the Bag model. As
in the case of the Diquark and constituent quarks models of the nucleon, the quarks in the NJL model are mostly free
apart from contact interactions. It is the relationship between the upper and lower components of a free Dirac spinor
that produce the sign of the Boer-Mulders function.

\bibliographystyle{plain}
\bibliography{proc}

\begin{thebibliography}{10}

\bibitem{Bacchetta:2004jz}
Alessandro Bacchetta, Umberto D'Alesio, Markus Diehl, and C.~Andy Miller.
\newblock {Single-spin asymmetries: The Trento conventions}.
\newblock {\em Phys. Rev.}, D70:117504, 2004.

\bibitem{Bacchetta:2003rz}
Alessandro Bacchetta, Andreas Schaefer, and Jian-Jun Yang.
\newblock {Sivers function in a spectator model with axial-vector diquarks}.
\newblock {\em Phys. Lett.}, B578:109--118, 2004.

\bibitem{Boer:2002ju}
Daniel Boer, Stanley~J. Brodsky, and Dae~Sung Hwang.
\newblock {Initial state interactions in the unpolarized Drell-Yan process}.
\newblock {\em Phys. Rev.}, D67:054003, 2003.

\bibitem{Brodsky:2002rv}
Stanley~J. Brodsky, Dae~Sung Hwang, and Ivan Schmidt.
\newblock {Initial-state interactions and single-spin asymmetries in Drell-Yan
  processes}.
\newblock {\em Nucl. Phys.}, B642:344--356, 2002.

\bibitem{Burkardt:2005hp}
Matthias Burkardt.
\newblock {Transverse deformation of parton distributions and transversity
  decomposition of angular momentum}.
\newblock {\em Phys. Rev.}, D72:094020, 2005.

\bibitem{Burkardt:2007xm}
Matthias Burkardt and Brian Hannafious.
\newblock {Are all Boer-Mulders functions alike?}
\newblock {\em Phys. Lett.}, B658:130--137, 2008.

\bibitem{Burkardt:2003je}
Matthias Burkardt and Dae~Sung Hwang.
\newblock {Sivers asymmetry and generalized parton distributions in impact
  parameter space}.
\newblock {\em Phys. Rev.}, D69:074032, 2004.

\bibitem{Diehl:2005jf}
M.~Diehl and Ph. Hagler.
\newblock Spin densities in the transverse plane and generalized transversity
  distributions.
\newblock {\em Eur. Phys. J.}, C44:87--101, 2005.

\bibitem{Gamberg:2003ey}
Leonard~P. Gamberg, Gary~R. Goldstein, and Karo~A. Oganessyan.
\newblock {Novel transversity properties in semi-inclusive deep inelastic
  scattering}.
\newblock {\em Phys. Rev.}, D67:071504, 2003.

\bibitem{Gockeler:2006zu}
M.~Gockeler et~al.
\newblock {Transverse spin structure of the nucleon from lattice QCD
  simulations}.
\newblock {\em Phys. Rev. Lett.}, 98:222001, 2007.

\bibitem{Pasquini:2007xz}
B.~Pasquini and S.~Boffi.
\newblock {Nucleon spin densities in a light-front constituent quark model}.
\newblock {\em Phys. Lett.}, B653:23--28, 2007.

\bibitem{Radici:2007vc}
Marco Radici, Francesco Conti, Alessandro Bacchetta, and Andrea Bianconi.
\newblock {Nucleon Spin Structure with hadronic collisions at COMPASS}.
\newblock 2007.

\end{thebibliography}

\end{document}